\documentclass[preprint2]{aastex}

\usepackage{natbib}
\usepackage{wasysym}
\usepackage{txfonts}
\def\W13{\mbox{${\rm W}_{\rm 13}$}}

\def\nH2{\mbox{${\rm n}(\HH$)}}
\def\enH2{\mbox{$n_{(\HH$)}}}

\def\pccc{~{\rm cm}^{-3}} 
\def\cmm5{~{\rm cm}^{-5}} 
 
\def\pcc {~{\rm cm}^{-2}}

\def\Tsub#1 {\mbox{$T_#1$}}
\def\TK  {\Tsub K }
\def\TB  {\Tsub B }

\def\fH2{\mbox{f$_\HH$}}
\def\mfH2{\mbox{$<{\rm f}_\HH>$}}

\def\p{\mbox{$^+$}}

\def\h13cop{\mbox{{H$^{13}$CO\p}}}

\def\c3h2{\mbox{C$_3$H$_2$}}
 
 \def\R0{R$_0$}
\def\G0{\mbox{G$_0$}}

\def\ddeg{{}^\circ\kern-.1em}

\def\kms{\mbox{km\,s$^{-1}$}}
\def\ps{\mbox{~s$^{-1}$}}

\def\E#1 {$10^{#1}$}
\def\E#1 {E{#1}}
\def\P#1,{$\nH2\TK~=~#1\times~10^4\pccc$~K}
\def\ec#1,#2,#3,{#1\,(#2)\E{#3}}

\def\H3{\mbox{H$_3$}}

\def\RH2{\mbox{R$_{\rm G}$}}
\def\GH2{\mbox{$\Gamma_{\HH}$}}
\def\g13{\mbox{g$_{13}$}}

\def\kHeH2{\mbox{$k_{ He-\HH}$}}

\def\tim#1,#2{\mbox{{$#1\times10^{#2}$}}}

\newcommand{\emm}[1]{\ensuremath{#1}}   
\newcommand{\emr}[1]{\emm{\mathrm{#1}}} 

\newcommand{\hcop}{\emr{HCO^+}} 
 
\newcommand{\HH}{\emr{{\rm H}_2}}
\newcommand{\cotw}{\emr{^{12}CO}}

\shorttitle{Excitation in bulk}
\shortauthors{H. S. Liszt and J. Pety}

\begin{document}



\title{The detectability of mm-wave molecular rotational transitions}


\author{Harvey S. Liszt}
\affil{National Radio Astronomy Observatory \\
            520 Edgemont Road,
           Charlottesville, VA,
           22903-2475}
\email{hliszt@nrao.edu}
\and
\author{Jerome Pety}
\affil{ Institut de Radioastronomie Millim\'etrique,
        300 Rue de la Piscine, F-38406 Saint Martin d'H\`eres, France \\ 
 Observatoire de Paris (CNRS UMR 8112), 61 av. de l'Observatoire, 75014, 
   Paris, France
\email{pety@iram.fr}}



\begin{abstract}

 Elaborating on a formalism that was first expressed some 40 years ago,
we consider the brightness of low-lying mm-wave rotational lines of
strongly polar molecules at the threshold of detectability.  We derive a
simple expression relating the brightness to the line of sight integral
of the  product of the total gas and molecular number densities and
a suitably-defined temperature-dependent excitation rate into the upper
level of the transition.   Detectability  of a line is contingent only on
the ability of a molecule to channel enough of the ambient thermal energy into
the line and the excitation can be computed in bulk by summing over rates
without solving the multi-level rate equations or computing optical
depths and excitation temperatures.
Results for \hcop, HNC and CS are compared with escape-probability
solutions of the rate equations using closed-form expressions for the
expected range of validity of our {\it ansatz}, with the result that
gas number densities as high as $10^4 \pccc$ or optical depths as high as
100 can be accommodated in some cases.  For densities below
a well-defined upper bound, the range of validity of the discussion
can be cast as an upper bound on the line brightness which is 0.3 K for
the J=1-0 lines and 0.8 - 1.7 K for the J=2-1 lines of these species.
The discussion casts new light on
interpretation of line brightnesses under conditions of weak
excitation, simplifies derivation of physical parameters and
eliminates the need to construct grids of numerical solutions of
the rate equations.

\end{abstract}



\section{Introduction}

MM-wave rotational transitions in the 3mm band are the workhorses
of interstellar chemistry and molecular line radioastronomy.  Their analysis
usually consists of a combined excitation/radiative transfer calculation
that derives a set of rotational excitation temperatures, optical depths and 
emergent line intensities as for instance outlined in the widely used
large velocity gradient (LVG) approximation by \cite{GolKwa74} and
embodied in various standard tools like RADEX and the Meudon PDR code
\citep[][respectively]{RADEX,LevLeP+12}.  A comprehensive review of
the constituent steps is given by \cite{ManShi15}.

Because the analysis is cast in terms of excitation temperatures and optical
depths, discussion  of the practical observability of a species or transition
is often expressed in terms of the so-called critical density that is sufficient to
excite the transition above the cosmic microwave background (which peaks in
the 3mm band).  Although somewhat loosely defined, a typical use
of the critical density equates the downward collision rate of the upper
level of a transition to its spontaneous emission rate.  \cite{Shi15}
defines the optically-thin critical density by equating the spontaneous
emission rate to the total rate of excitation, up and down, out of a
given level.  Critical densities are often
quite large, eg $\ga 10^{5}\pccc$ and much higher than the densities
that are derived from detailed analysis of quite bright lines.  
\cite{Shi15} shows how the estimates of the required density can be decreased 
when other considerations, for instance radiative trapping, are included.


Here we develop an alternative approach to calculating the brightness of
molecular transitions when the density is far below the critical density, 
 an approach that was first introduced by \cite{Pen75} and subsequently
elaborated for the J=1-0 transition of HCN by \cite{LinGol+77}.  
Some aspects of this  methodology have been cited in the interpretation of 
sub-mm observations of water \citep{WanPag+91,SneHow+00} and the two-level 
fine-structure excitation of C\p\ \citep{GolLan+12} but they are much 
less commonly recognized in the context of mm-wave transitions.  
In any case, the formalism is broader, more powerful and more generally 
useful than is apparent from  these earlier references, as we 
hope will be made clear in this work. 

Given that the critical density is so high, this approach is applicable even 
to fairly dense gas under conditions of appreciable optical depth.
The underlying physics is simply that when collisional excitation is weak, 
collisional de-excitation is also weak:  Downward collisions are rare 
compared to spontaneous and stimulated emission, so energy put into the 
rotation ladder by upward collisions eventually emerges from the gas 
even if it is repeatedly absorbed and scattered, not only  when
the medium is  optically thin.  The last remnant of
energy injected by an upward collision from the J=0 level
eventually emerges in the J=1-0 line.  Calculating the brightness of a line
consists merely of determining the rate at which collisions are putting 
energy into the line without recourse to solution of the rate equations
for the level populations, the optical depth, and indeed, without detailed
knowledge of the spontaneous emission rate.

The ultimate point is that for any transition or species, no matter how low 
the ambient number density of collision partners, there is a molecular column 
density 
that will produce a given brightness.  {\it A priori} there is really no 
minimum density needed to produce a detectable line, only a required number 
density-column density product that we define as the molecular 
emission measure because (as we show) it is the line of sight integral
of the product of two number densities.  The constant of proportionality 
between the emission measure and line brightness depends on the collision
rate and to a minor extent on the explicit molecular structure,
i.e., the  rate coefficients for collisions with ambient
particles and the  wavelength at which a transition occurs, 
but not the optical depth or spontaneous emission rate.

The structure of this work is as follows.  In Section 2 we 
 re-derive the expression for the emergent brightness of weakly 
excited lines in the framework of the underlying physics of ``bulk excitation'' 
and we discuss collisional excitation  of \hcop, HNC and CS by 
\HH, He and electrons.  In Section 3 we discuss the bulk excitation 
of \hcop, HNC and CS in the framework of the escape probability 
formalism underlying the LVG approximation and we show that LVG and 
bulk excitation calculations yield the same results when rather broad 
limits (which we derive) on the number density and line brightness 
are observed.  Section 4 gives a brief discussion of some implications
and limitations and Section 5 is a summary.

\begin{figure*}
 \includegraphics[height=7.8cm]{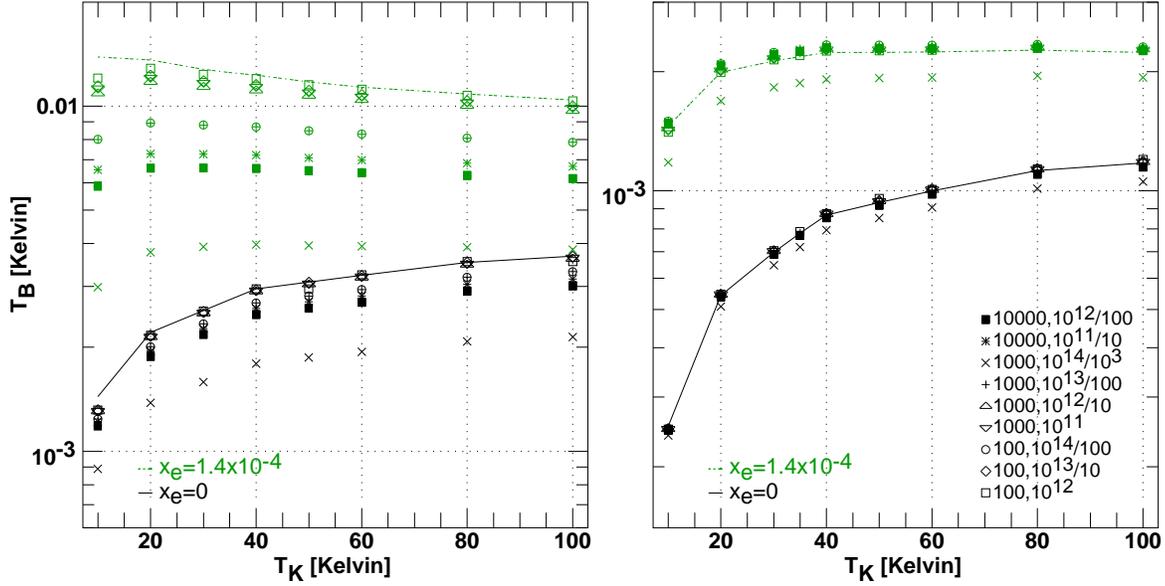}
  \caption[]{
Comparison with LVG calculations for \hcop\ J=1-0 at left and J=2-1 at 
right.  Black (solid) and green (dash-dot) lines show calculations using 
the bulk excitation framework discussed in Section 2 with 
n(H) = $100 \pccc$ and dN(\hcop)/dv = $10^{12}\pcc$ (\kms)$^{-1}$, 
for gases with a 
molecular hydrogen fraction of unity (n(\HH) = n(H)/2) and electron 
fractions x$_e$ = n(e)/n(H) = 0 and $1.4\times 10^{-4}$ respectively.  
LVG calculations of the brightness were performed at each value of 
x$_e$ with  parameter combinations n(H), N(\hcop) indicated at the lower 
right in the right-hand panel and with symbols as indicated there. 
Each n(H), N(\hcop) 
parameter combination is represented by a black and a green symbol.
Results of the LVG calculations  
were scaled by $10^{14}$/(n(H)dN(\hcop)/dv) to test the linear
dependence that is expected in the limit of weak excitation.  For example,
the brightnesses of the LVG calculations with n(H) $= 10^4 \pccc$, 
dN(\hcop)/dv $= 10^{12}\pcc$ represented by the solid rectangles were divided
by a factor 100 as indicated in the figure legends. 
}
\end{figure*}

\begin{figure*}
 \includegraphics[height=7.8cm]{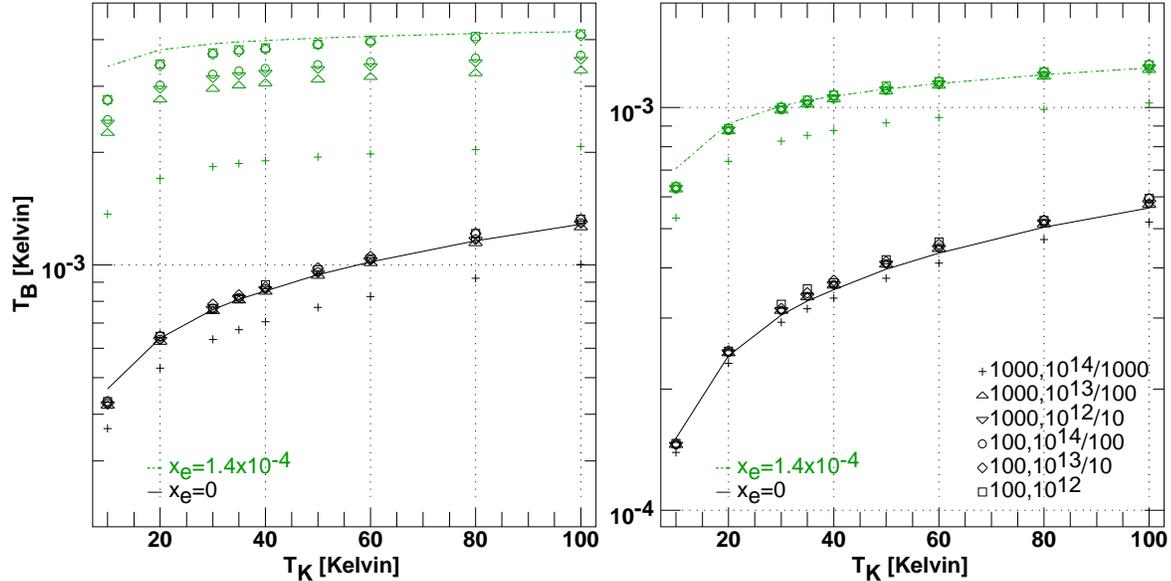}
  \caption[]{As in Figure 1, but for CS.}
\end{figure*}

\section{Weakly-excited lines}

\subsection{Basic concepts}

The approach taken here is a perturbation approximation where the
equilibrium condition throughout the molecular rotational energy
ladder is pure radiative equilibrium with the cosmic microwave background
(CMB) at 2.725 K, minutely  disturbed by collisions with ambient particles
at the local kinetic temperature \TK.

Consider the canonical two-level atom with  a transition at frequency 
$\nu$, immersed in a gas of total density of H-nuclei n(H) and kinetic 
temperature \TK.  The lower and upper levels are labelled {\it l} and {\it u}, 
respectively, at energies E$_l$ and E$_u$ above the ground level, having 
statistical weights g$_l$ and g$_u$.  The number densties of
molecules in the lower and upper levels are n$_l$ and n$_u$ and 
the levels are connected by collisions having 
rate coefficients $\gamma_{lu}$ and $\gamma_{ul}$ (units of cm$^3~\ps$) 
where $\gamma_{ul}/\gamma_{lu}$ = g$_l$/g$_u \exp(h\nu/k\TK)$.
The upward and downward collision rates (units of $\ps$)
are C$_{lu}$ = n(H)$\gamma_{lu}$ and C$_{ul}$ = n(H)$\gamma_{ul}$; 
the rate coefficients $\gamma$ are normalized to reflect this definition
as discussed below.

The levels are also connected by the spontaneous emission rate of the upper level,
A$_{ul}$, and by transitions induced by the cosmic microwave background at 
temperature T$_{cmb}$ =2.725 K. The exact solution for the level populations, 
including the radiation field of the cosmic microwave background is
$$ n_u/n_l = (g_u/g_l) \frac{[p_\nu(T_{cmb}) + \exp(-h\nu/k\TK) C_{ul}/A_{ul}]}
  {[p_\nu(T_{cmb})+1+ C_{ul}/A_{ul}]} \eqno(1)$$
where $p_\nu(x) = 1/(\exp(h\nu/kx)-1)$ and $1+p_\nu(x) = 1/(1-\exp(-h\nu/kx))$ .  

For low-lying rotation transitions in the 3mm band, h$\nu$/k = 4.3 K 
($\nu$/100 GHz) and g$_l$/g$_u < 1$, so, at typical kinetic temperatures 
$>$ 4 K it follows that $\gamma_{ul} < \gamma_{lu}$ and
${\rm C}_{ul} < {\rm C}_{lu}$:  If collisional
excitation is weak, C$_{lu} <<$ A$_{ul}$, collisional de-excitation
is even weaker.  Thus in the limit of weak excitation, for the 
transitions considered here with relatively small values of 
$h\nu/k$,  we will also have $C_{ul} < C_{lu} << A_{ul}$ and the 
excess population in the level $u$ over that which would obtain 
in radiative equilibrium with the  cosmic microwave background is 
$$\Delta n_u = \frac{n_l (C_{lu}/A_{ul})}{1+p_\nu(T_{cmb})}. \eqno(2)$$
This is obtained by differencing Eq. 1 evaluated with and without
collisions and neglecting the collisional term in the denominator:
note the change from ${\rm C}_{ul}$ in Eq. 1 to ${\rm C}_{lu}$ in
Eq. 2 and succeeding equations. Since $\Delta n_l = -\Delta n_u$,
the fractional shift of population out of the lower level, away 
from the radiative equilibrium established by the cosmic background,
$|\Delta n_l/n_l| = (C_{lu}/A_{ul}) /(1+p_\nu(T_{cmb}))$, is very
small.

\begin{figure}
 \includegraphics[height=5.35cm]{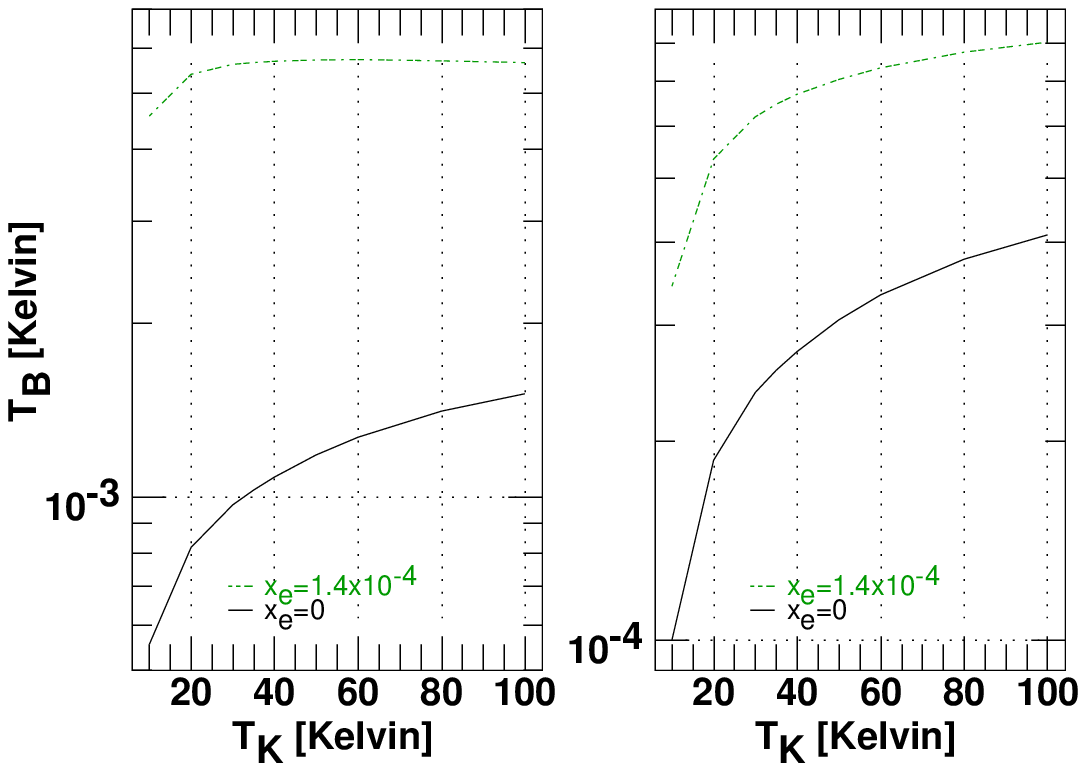}
  \caption[]{Excitation curves for n(H)$=100\pccc$, 
dN(HNC)/dv $=10^{12}\pcc$(\kms)$^{-1}$ as in Figures 1 and 2, for HNC J=1-0 
(left) and J=2-1.}
\end{figure}

The collision-induced volume emissivity of the line
$\epsilon_u$ (erg $\pccc$\ps\ sr$^{-1}$) is

$$ \epsilon_u = \Delta n_u (h\nu/4\pi) A_{ul} 
 = \frac{(h\nu/4\pi) n_l C_{lu}}{1+p_\nu(T_{cmb})} \eqno(3)$$
independent of A$_{ul}$.  This is the origin of the approach taken here:
in the limit of weak collisional excitation (WCE),
spectral lines are merely conduits for the energy that is
deposited into them and the energy is radiated at the same rate that it
is injected, not at rates determined by the A-coefficients. 
Independence from the spontaneous emission coefficient that
is the main determinant of the critical density has clear implications
for the limiting spatial extent of lines of different chemical species.

The simultaneous balancing of the cosmic bacground
radiation provides a slight drag on the efficiency of the collisional energy
transfer into the line, ie 1+$p_\nu(T_{cmb}) = 1.26$ at 90 GHz. 

\subsection{Summing to a total rate and column density}

In the linear molecules considered here, all upward transitions out 
of lower-lying levels J $<$ J$_u$ inject energy into the 
$J_u$ - $J_l$ line.  Thus for the 1-0 line, all rates $C_{0,J^\prime}$
for the 0-1, 0-2, etc. transitions are summed; for the J=2-1 line, rates for 
the transitions 0-2, 0-3 ... 0-n and 1-2, 1-3 ... 1-n.

Specifically, we define an excitation rate into the level $J_u$ as

$$ C_u = \Sigma_{j <j_u}\Sigma_{j^\prime \ge j_u} f_j 
 C_{j,j^\prime}/(1+p_\nu(T_{cmb})),  \eqno(4) $$
where the CMB correction is included on a term-by-term basis
as a modification of the collision rates $C_{J,J^\prime}$\ and 
C$_u$ is normalized so that the effective excitation 
rate coefficient into the level u, $\gamma_u$, is given by

$$ C_u = n(H) \gamma_u.  \eqno(5)$$
The fraction of molecules in the $l$-th level is calculated in
equilibrium with the cosmic microwave background

$$f_l = g_l \exp(-E_l/kT_{cmb})/Q(T_{cmb})  \eqno(6)$$
and the partition function Q(T$_{cmb}$) is calculated as 

$$Q(T_{cmb})= \Sigma_l  g_l \exp(-E_l/kT_{cmb})  \eqno(7)$$
with $T_{cmb}$ = 2.725 K.  The total number density of molecules 
summed over all levels is just $n_l/f_l$.  Some 60\% of the total 
rotational population is in the J=0 level for \hcop\ and HNC in 
radiative equilibrium with the CMB,  or 40\% for CS.  The only
temperature that appears explicitly in these equations is that of the cosmic
background, and the kinetic temperature dependence is entirely 
contained in the rate coefficients $\gamma_u$.
  
Finally, we  define a density n(H)$_{max}$ for the level u as a fraction
q (q $<<1$) of the density at which the total excitation rate into the level 
u equals its spontaneous emission rate, 

$$  n(H)_{max} \equiv q A_{ul}/\gamma_u, \eqno(8) $$
 independent of the optical depth.
Limiting the density in this way is necessary to justify
the approximations leading to Eq. 3, but the importance of the requirement that 
 n(H) $\le$ n(H)$_{max}$ should not be understood merely as a limit on the
excitation temperature:  It is also necessary to allow energy injected
into the line by collisions to escape the medium without being re-absorbed
back into the thermal reservoir of the gas by collisional de-excitation.
 The weak collisional excitation (WCE) approximation with a fixed 
proportionality between line brightness and the emission measure only applies  
when n(H) $\le$ n(H)$_{max}$.  This is discussed in more detail in Section 3 
where a value q = 1/8 is derived by considering the relationship of the line
brightness and the emission measure in the limits of high density and 
implicit optical depth.  When n(H) $>$ n(H)$_{max}$ there may still 
be a regime where for suitably faint lines the brightness is linearly 
proportional to column density (Figure 3 of \cite{LinGol+77}), but 
the ratio of brightness to emission measure decreases because 
progressively fewer excitations directly result in observable photons. 

Our n(H)$_{max}$ is different from the usual critical density because it is 
defined in terms of upward rates and because it sums over collisions that 
do not directly involve the level in question (for instance 
$\gamma_1$ includes 0-1, 0-2, 0-3, etc collisions).  The values
of A$_{ul}/\gamma_u$ are quite similar 
to the optically-thin critical densities discussed by \cite{Shi15}
but our n(H)$_{max}$ are smaller by the factor q.    
Values of n(H)$_{max}$ are given in Tables 2 and 3 for the case q = 1/8 
that is discussed in Section 3.  For excitation by neutral particles
(Table 2), n(H)$_{max}$ = $5 - 10 \times 10^3\pccc$ 
for the J=1-0 lines of \hcop and HNC, and about 10 times higher for 
their J=2-1 transitions. Values of n(H)$_{max}$ for CS are about 
5 times smaller for both lines.  When electron excitation dominates
for x$_e = 1.4\times 10^{-4}$ (Table 3) the limiting densities 
 n(H)$_max$ are on the whole lower by a factor of order 10.

Importantly, note in Tables 2 and 3 that Einstein A-coefficients increase 
much more rapidly with J$_u$ than do the collision rates and the limiting 
densities n(H)$_{max}$ increase with J$_u$, quite rapidly  at the bottom of 
the rotation ladder.  Thus if the weak-excitation case applies 
to any transition, it is even more valid for yet higher-lying lines.

\subsection{Observable line brightness}

The total specific energy flux W$_u$ (erg cm$^{-2}~\ps$ sr$^{-1}$) from a bulk medium 
is the line of sight integral of the emissivity $\epsilon_u$  , ie

$$W_u = \int \epsilon_u dL  = (h\nu/4\pi)
 \int n(H) \gamma_{u} n(Y) dL  \eqno(9) $$
where $\gamma_{u}$ is kept inside the integral to account for variations
in the kinetic temperature: The integral is defined and discussed here
as the  molecular emission measure. The radiative transfer of reabsorption by 
intervening material along the line of sight is ignored here with the 
understanding that the range of validity of such an assumption will be
derived.  The physical basis for the assumption is the limitation
to densities below n(H)$_{max}$ so that scattered photons are
not re-absorbed back into the bath of thermal energy by downward collisions.


If the relative abundance of species Y, X(Y) = n(Y)/n(H)  is constant, 
the emitted energy W$_u \propto$ X(Y) $\int  n(H)^2$ dL is 
heavily weighted 
to regions of higher density and strongly influenced by clumping.  
If the density n(H) and relative abundance 
X(Y) are constant, W $\propto$ n(H) N(Y) where N(Y) is the total
column density of species Y (this is the analog of the LVG calculation
described in Section 3).  For any density n(H) $\le$ n(H)$_{max}$ 
there is a column density N(Y) that will produce a given output W. 
The observability of a line is determined by the  molecular emission measure,
essentially the product of the density and column density.

In terms of observables, the integrated line brightness temperature 
\TB\ above the black-body background is determined by

$$ \int 2 k \TB/\lambda^2 d\nu= W_u  \eqno(10)$$
or, in velocity units (\kms) using Eq. 9

$$\int \TB d{\rm v} = (\lambda^2/8\times 10^5 \pi)(hc/k)
  \int n(H)  \gamma_u n(Y) dL   \eqno(11) $$

\subsection{Actual calculation of the collision rate}

The rate coefficients $\gamma_u$ are comprised of weighted contributions 
from collisions with helium, molecular hydrogen and electrons, ie

$$ \gamma_u = 0.0875 \gamma_u({\rm He}) + x_e \gamma_u(e) + (\fH2/2) 
\gamma_u(\HH)  \eqno(12)$$ 
with n(He)/n(H) = 0.0875 \citep{Bal06}, x$_e$ = n(e)/n(H) and \fH2\ = 2n(\HH)/n(H) 
(maintaining the usual definition of the latter).  For the cases where
all the hydrogen is in \HH, n(\HH) = n(H)/2.  We ignore
collisions with atomic hydrogen that is generally ineffective at exciting
molecules (and consider gas in which all the hydrogen is molecular) but include 
collisions with electrons having a proportion x$_e$ = n(e)/n(H).  With the 
exception of CO to which the WCE limit does not apply over an interesting
range of n(H) and/or N(CO) (see Section 4.1), collisions 
with electrons will usually dominate the excitation when
CO does not bear the majority of the gas phase carbon, see \cite{Lis12}.
References to the collision rate coefficients used here are given in Table 1
and the dominance of electron excitation is apparent when collision rates
are tabulated (see Appendix A).

\subsubsection{Excitation by electrons}

For strongly-polar species having permanent dipole moments $\mu > 0.5$ 
Debye, excitation by electrons has a strongly dipole character 
($\Delta J = +1$) and rate constants are well represented in separate 
closed forms for molecular ions \citep{DicFlo81,BhaBha+81,NeuDal89} and 
neutrals \citep{DicPhi+77}.  However, more accurate rates for e-\hcop\ and e-HNC 
collisions have been calculated by \cite{FauTen+07} and \cite{FauVar+07}, 
respectively and for CS by \cite{VarFau+10}.  These have 
the additional virtue that they include the smaller terms with $|\Delta J| > 1$ 
that are important for determining line brightness ratios over the rotation ladder 
in a single species.

\begin{table*}
\caption[]{References to collision rate coefficients}
{
\begin{tabular}{lccc}

\hline
Species & \HH & He & electrons \\
\hline
\hcop & \cite{Flo99}$^1$ & scaled \HH$^2$ & \cite{FauTen+07} \\
HNC & \cite{DumKlo+11} & \cite{DumFau+10} & \cite{FauVar+07} \\
CS & scaled He & \cite{LiqSpi+06} & \cite{VarFau+10} \\
\hline
$^1$ See also \cite{YazBen+14} \\
$^2$ See also \cite{BufDor+09} \\
\end{tabular}}
\end{table*}

\subsubsection{Excitation by  helium}

Recent calculations for He-molecule collisions are available for HNC and CS, 
as noted in Table 1.  For \hcop\ there is a recent reference by \cite{BufDor+09}
but the appendix containing the tables of rate coefficients is missing in
the article online so we used the \HH-\hcop\ rates scaled downward
by a factor 1.4, which is the inverse of the common practice when collisions
with He must be substituted for those with \HH.  Examination of the few rates
that were tabulated by \cite{BufDor+09} for comparison with earlier
results showed that the error introduced by scaling the \HH\ rates will
be very small, especially given the He abundance.

\subsubsection{Excitation by molecular hydrogen}

These are noted in Table 1.  For HNC both ortho (odd-J) and para \HH\ collision
rates are considered separately and equilibrium between the ortho and para \HH\
populations at the kinetic temperature was assumed:  this is very likely to be
correct for the case that the electron fraction is appreciable, due to the
population of protons. For \hcop\ we used the para-\HH\ rates of \cite{Flo99}
as in \cite{Lis12}: a more recent calculation \citep{YazBen+14} differs 
typically by 10\% or less. For CS, unfortunately, the only recent 
 collision rate calculations with neutral particles are those for 
helium \citep{LiqSpi+06} but these are also used in RADEX (with appropriate
 upward scaling  by a factor 1.4) so there is at least a basis 
for comparison.

\subsection{Overall behaviour}

The individual terms contributing to $\gamma_1$  are
shown in the Tables  in Appendix A for \hcop, HNC and CS excited by 
He and \HH\ with and without electrons.  For excitation by neutral
particles, J=0-1 transitions dominate the excitation into the J=1 level at 
low temperature but only one half - one-third of the total rate is due 
to direct excitations from the ground state at/above 20 K.  This
is the origin of the temperature dependence of the line brightnesses
shown in Figures 1-3.   For a gas 
with x$_e = 1.4\times 10^{-4}$, the collision rate constant per hydrogen
is 10-20 times larger but there is little temperature dependence
of the collision scheme  and 
direct excitations into the upper level of a transition dominate.

\subsection{Scaling with the electron and \HH\ fractions}

When electron collisions dominate at x$_e = 1.4\times 10^{-4}$ the results 
here can be considered to be independent of \fH2\ because the density of 
electrons n(e) = 0.00014 n(H) has been assumed to be equal to the total 
density of gas-phase carbon.  When electron excitation dominates 
(see Tables 5 and 6) the implied total density n(H) can be scaled to
other electron fractions x$_e$ by keeping the electron density 
n(e) = x$_e$n(H) constant.

For simplicity the calculations here present only the case \fH2 = 1
and ignore collisional excitation by atomic hydrogen, which is
generally not considered to be important enough to merit calculation
of the excitation rate.  Note that the very high rates for
excitation of CO by atomic hydrogen calculated by \cite{BalYan+02}
and discussed by \cite{Lis06} were subsequently refuted by 
\cite{SheYan+07} and \cite{YanSta+13}.

When \HH\ excitation dominates, an approximate scaling may be derived
by accounting for the fact that n(\HH) =  n(H)\fH2/2 and that
the rate constants for helium are generally considered to be
smaller than those for \HH\ by a factor 1.4 following the usual 
scaling by the thermal speed.  This implies a  scaling with
\fH2\ such that n(H)(0.125+\fH2)/1.125 $\approx$ constant.

\section{Validation}

The physics discussed here is unexceptionable but as a check on the results 
and to ascertain the limits of validity of the formalism in opaque media
we compare with the 
results of LVG calculations \citep{GolKwa74}.  An LVG calculation amounts to  
an evaluation of Eq. 11 at constant density and temperature.  Comparison with 
this widely-used approximation to the radiative transfer problem provides a 
check on the normalization of the brightness and the effects of radiative
trapping which to some degree circumscribe the range over which the 
excitation can be considered to be weak and the medium may be regarded
as being transparent.

In the LVG approximation described by \cite{GolKwa74}, the Einstein 
spontaneous emission coefficients A$_{ul}$ are replaced by 
A$_{ul}\beta_{lu}$ where $\beta_{lu} = (1-\exp(-\tau_{lu}))/\tau_{lu}$ 
is the photon escape probability derived by \cite{Cas70} and $\tau_{lu}$ is 
the line optical depth.  In the limit of high optical depth, $\beta_{lu} 
\rightarrow 1/\tau_{lu}$ and A$_{ul}\beta_{lu} \rightarrow A_{ul}/\tau_{ul}$,
and  to account for cases of high optical depth we rewrite Eq. 8 as 

$$ C_{u}\tau_{lu}/A_{ul} = n(H) \gamma_{u}\tau_{lu}/A_{ul} \le q. \eqno(13)$$ 
However, $\tau_{lu}/A_{ul}$ is independent of A$_{ul}$ and after expressing 
$\tau_{lu}$ in terms of dN(Y)/dv and A$_{ul}$ {\it etc.} using standard 
formulae relating column density and optical depth \citep{Spi78} we can
recast Eq. 13 as a limit on the density-column density product, 

$$ n(H) \frac{dN(Y)}{d{\rm V}} \le q \frac{8\times10^5 \pi Q(T_{cmb}) 
\exp(E_l/kT_{cmb})} {g_u\lambda^3  \gamma_{u}} \eqno(14)  $$
with velocity expressed in \kms.  There is only a slow temperature dependence
over the range 10 - 80 K and limiting values for \TK\ = 30 K
are shown in Table 4 for q = 1/8 with $x_e=0$ and $x_e = 1.4\times 10^{-4}$.

We chose q = 1/8 after running many models because it limits the divergence
from the LVG calculations to 20\% or less when n(H) $\la$ n(H)$_{max}$.  For the
species discussed here, using q = 1/8 to limit the upward rate C$_u$ is equivalent 
to C$_{10}$/A$_{10} \la 1/20$ at 10 K or C$_{10}$/A$_{10} \la 1/50$ at/above
30 K which certainly justifies the approximations that were assumed to derive 
the basic equations.

Using Eq. 11, Eq. 14 can be rewritten as an upper limit on the
observed line brightness in remarkably compact form, namely


$$ T_B \le q \frac{h\nu}{k} \frac{Q(T_{cmb})\exp(E_l/kT_{cmb})}{g_u}, \eqno(15) $$
showing that the applicability of the WCE approximation can be understood 
solely in terms of the line brightness, independent of the means of excitation 
or the underlying physical conditions, given only that n(H) $\le$ n(H)$_{max}$. 
A similar conclusion can be found in the text of \cite{LinGol+77} following
their Equation 15, but without the explicit caveat on the density.
Values of the limiting brightness temperature are given in Table 4.  
These are 0.3 K for the J=1-0 lines and 0.8 - 1.7 K for J=2-1.

\subsection{Computational results}

Figures 1 and 2 show direct comparisons between the bulk excitation results 
and those of full LVG calculations, for \hcop\ and CS respectively, with
the J=1-0 lines at left and J=2-1 at right in each case.  Results are shown 
for  x$_e = 0 $ in black (these lay lower) and for $x_e = 1.4\times 10^{-4}$ 
in green.  Eq. 12 was evaluated at n(H) $= 100\pccc$ and molecular column 
densities dN(mol)/dv $=10^{12}\pcc (\kms)^{-1}$, ie.
n(H)dN(mol)/dv $= 10^{14}$cm$^{-5} (\kms)^{-1}$.  
The LVG calculations were carried out at various density, column density products 
as shown in the panels of the figures and their results were normalized to a 
product of $10^{14} {\rm cm}^{-5} (\kms)^{-1}$: for instance, the LVG brightnesses 
computed for n(H) = $10^3 \pccc$, dN(mol)/dv $= 10^{13}\pcc(\kms)^{-1}$ were 
divided by 100.  The actual line
brightnesses represented in the panels vary by factors up to almost $10^{3}$ and the 
optical depths range up to almost 100 for the J=1-0 line of \hcop\ while
retaining a high degree of linear proportionality to the density-column density
product within the weak-excitation regime as described in the previous Section.  

\begin{table*}
\caption[]{Limiting density n(H)$_{max}$ for q = 1/8 and $x_e = 0$}
{
\begin{tabular}{ccccccccc}
\hline
Species & $J_u-J_l$ &  5 K  &  10 K   &  20 K   &  30 K   &  40 K   &  60 K  &  80 K  \\
       & &&& &$\pccc$  & & & \\
\hline
\hcop& 1-0 & 1.79E+04 & 1.39E+04 & 9.01E+03 & 7.77E+03 & 6.72E+03 & 6.13E+03 & 5.63E+03 \\
& 2-1 & 3.40E+05 & 1.89E+05 & 8.77E+04 & 6.83E+04 & 5.49E+04 & 4.77E+04 & 4.23E+04 \\
HNC& 1-0 & 4.00E+04 & 2.22E+04 & 1.51E+04 & 1.27E+04 & 1.14E+04 & 9.71E+03 & 8.76E+03 \\
& 2-1 & 8.93E+05 & 2.94E+05 & 1.58E+05 & 1.25E+05 & 1.08E+05 & 8.89E+04 & 7.84E+04 \\
CS& 1-0 & 1.22E+04 & 3.64E+03 & 2.66E+03 & 2.23E+03 & 1.99E+03 & 1.66E+03 & 1.46E+03 \\
& 2-1 & 1.06E+05 & 2.45E+04 & 1.53E+04 & 1.21E+04 & 1.05E+04 & 8.48E+03 & 7.33E+03 \\
\hline
\end{tabular}}
\\
\end{table*}

\begin{table*}
\caption[]{Limiting density n(H)$_{max}$ for q = 1/8 and $x_e = 1.4\times 10^{-4}$}
{
\begin{tabular}{ccccccccc}
\hline
Species & $J_u-J_l$ &  5 K  &  10 K   &  20 K   &  30 K   &  40 K   &  60 K  &  80 K  \\
       & &&& $\pccc$  & & & & \\
\hline
\hcop& 1-0 & 1.68E+03 & 1.43E+03 & 1.45E+03 & 1.55E+03 & 1.61E+03 & 1.74E+03 & 1.83E+03 \\
& 2-1 & 6.17E+04 & 3.27E+04 & 2.39E+04 & 2.23E+04 & 2.12E+04 & 2.12E+04 & 2.10E+04 \\
HNC& 1-0 & 3.96E+03 & 2.70E+03 & 2.28E+03 & 2.19E+03 & 2.17E+03 & 2.15E+03 & 2.16E+03 \\
& 2-1 & 2.07E+05 & 8.64E+04 & 5.55E+04 & 4.79E+04 & 4.43E+04 & 4.04E+04 & 3.82E+04 \\
CS& 1-0 & 6.64E+02 & 4.99E+02 & 4.50E+02 & 4.34E+02 & 4.25E+02 & 4.15E+02 & 4.08E+02 \\
& 2-1 & 9.48E+03 & 5.22E+03 & 4.04E+03 & 3.66E+03 & 3.46E+03 & 3.21E+03 & 3.06E+03 \\
\hline
\end{tabular}}
\\
\end{table*}


\subsubsection{\hcop}
 
Figure 1 shows computational results for the J=1-0 and J=2-1 lines of \hcop\ at 
left and right, respectively.   The two lines have optical depth of unity  at 
dN(\hcop)/dv = 1.1 and 2.2 $\times 10^{12}\pcc$ (\kms)$^{-1}$, respectively.

Without electrons, the entries in Table 4 predict that the
bulk excitation and LVG calculations should diverge for 
n(H)dN(\hcop)/dv $\ga 2\times 10^{16}$cm$^{-5} (\kms)^{-1}$ at low density
and this is manifested in the calculations: with  n(H) $= 10^3 \pccc$, 
the deviation from the LVG calculation increases from 7\% at 
dN(\hcop)/dv = $10^{13}\pcc (\kms)^{-1}$ where \TB\ $\la$ 0.33K to 40\% at 
dN(\hcop)/dv = $10^{14}\pcc (\kms)^{-1}$ where \TB\ $\la$ 2K.
The density criterion is more important than a limit on the optical depth:
the deviation is larger for n(H)$ = 10^4 \pccc$, dN(\hcop)/dv 
$= 10^{12}\pcc(\kms)^{-1}$, where the optical depth is 0.9, than 
for  n(H) $= 10^2 \pcc$, dN(\hcop)/dv $= 10^{14}\pcc~(\kms)^{-1}$ where
it is 90.  The calculations also demonstrate the dependence on the
density-column density product, ie the results coincide for 
n(H) $= 10^2 \pccc$, dN(\hcop)/dv  $= 10^{14}\pcc~(\kms)^{-1}$ and 
for n(H) $= 10^3 \pccc$, dN(\hcop)/dv $= 10^{13}\pcc$ (\kms)$^{-1}$.

Also as predicted, the range of validity is more limited in terms of
both density and the density-column density product when electrons
dominate the excitation while observing the same limit on the 
brightness.  The ranges of brightness and n(H)dN(mol)/dv over which the
bulk excitation and LVG results coincide are much wider for the 
J=2-1 line than for J=1-0.

\subsubsection{CS}

Figure 2 shows computational results for the J=1-0 and J=2-1 lines of CS at 
left and right, respectively.  The two lines have optical depth of unity  at 
dN(CS)/dv = 9.8 and 8.2 $\times 10^{12}\pcc$ (\kms)$^{-1}$, respectively.

CS ostensibly differs from \hcop\ in several notable ways even if the 
brightness temperature limits in Table 4 are comparable:  The CS J=1-0 
transition lies  much lower (49 GHz) where the CMB correction is 2 times 
larger and the lines require 4-8 times higher column density to achieve unit 
optical depth. Cross-sections for excitation by neutral particles are 
much smaller because CS is physically more compact (Tables 5 and 6).  
The limiting densities in Table 2 are 2-3 times lower for CS than for 
\hcop\ but the limiting density-column density products are larger by 
the same factor and  CS requires about an order of magnitude 
higher column density to emit the same amount of energy in 
the J=1-0 line, compared to \hcop.    In the end much of this 
is compensated in the conversion to brightness temperature 
and (like HNC) the J=1-0 lines of CS are 3-4 times weaker than 
\hcop\ for a given density-column density product.

As for \hcop, the LVG and bulk-excitation calculations diverge when the 
density-column density product or brightness exceed the limiting values 
shown in Table 4.

\subsubsection{HNC}

Figure 3 shows computational results for the J=1-0 and J=2-1 lines of HNC at 
left and right, respectively.  The two lines have optical depth of unity  at 
dN(HNC)/dv = 1.8 and 3.7 $\times 10^{12}\pcc$ (\kms)$^{-1}$ 
respectively.  The parameters for HNC are similar to those for \hcop\ 
and results are shown without comparison with LVG results.

\begin{table*}
\caption[]{Limiting brightness temperature and n(H) dN(mol)/dv$^1$ for q = 1/8}
{
\begin{tabular}{ccccc}
\hline
Species & $J_u-J_l$ & \TB & n(H)dN(mol)/dv (x$_e=0)$ & n(H)dN(mol)/dv (x$_e=1.4\times10^{-4})$ \\
        &           & Kelvins & cm$^{-5}$(\kms)$^{-1}$ &cm$^{-5}$(\kms)$^{-1}$\\
\hline
\hcop & 1-0 & 0.30 & 1.9E16  & 1.9E15\\   
& 2-1    & 1.72 & 4.4E17 & 6.2E16 \\
HNC & 1-0 & 0.30 & 4.1E16 & 3.6E15\\
& 2-1    & 1.77 & 1.0E18 & 1.5E17\\
CS & 1-0 & 0.26 &4.4E16 & 3.0E15 \\
& 2-1    & 0.75&4.6E17 & 4.4E16\\
\hline
\end{tabular}}
\\
$^1$ Evaluated at \TK\ = 30 K \\
\end{table*}

\subsection{Using bulk excitation to replace or scale LVG calculations}


Despite the elaborate nature  of the preceding disussion, using the calculations 
illustrated in Figures 1 - 3 is straightforward.  Given a brightness that obeys
the limits in Table 4, or an integrated brightness whose peak brightness 
obeys the limit, divide by the value in the appropriate curve, multiply by 
$10^{14}$ and that is n(H)dN(Y)/dv or n(H)N(Y) at the assumed temperature
and electron fraction.  Dependencies on the assumed parameters are clearly
manifest in ways that are not always apparent from interpreting the numerical
results of large grids of LVG model calcuations.  Conversely, knowing a priori
how LVG results will scale,  a single LVG calculation can replace a grid of
numerical solutions.


As an example, consider the observations of \hcop\ J=1-0 lines in absorption in 
diffuse and translucent clouds with measured N(\hcop) $\approx 10^{12}\pcc$ 
accompanied by weak emission with integrated brightnesses 
$\la 0.03$ K-\kms\ \citep{LucLis96}.  At \TK\ = 30 K with 
x$_e = 1.4\times10^{-4}$ the implied density-column density 
product from Figure 1 is n(H)N(\hcop) $= 2.2\times10^{14}\cmm5$, 
implying n(H) $\approx 220\pccc$. Without electrons the derived density 
ranges from ten times higher at \TK\ = 10 K to three times at \TK\ = 80K.  

Alternatively, consider the weak but ubiquitous emission from strongly-polar 
species that accompanies carbon monoxide emission in surveys of the 
inner Galactic plane with brightness about 2\% that of \cotw\ 
\citep{Lis95,HelBli97}.
The widespread nature of this emission was somewhat surprising given the 
conventional wisdom about the difficulties of exciting it, but a variety
of considerations led to the conclusion that it was arising in gas of 
rather moderate density.  At 1\%-2\% the brightness of CO, the observed 
brightnesses of \hcop, HCN, CS, etc overwhelmingly lie within the range 
of applicability of our discussion.
A reanalysis of these observations using recent values for the 
molecular abundances and excitation rates should lead to an improved
understanding of the structure of inner-Galaxy  molecular clouds.

\section{Other considerations}

\subsection{Extension to other species}

The calculation described here works well for molecules that have low-lying
transitions in the mm-wave regime where an explicit limit on the 
upward collision rates for low-lying levels also 
suitably limits the downward rates, guaranteeing weak collisional
excitation while allowing the free escape of energy in the lines.  
Moreover, most commonly-observed molecules have permanent dipole moments 
of about 1 Debye or more, making the spontaneous emission rates high enough to 
allow consideration of an interesting range of parameter space.
However many commonly-observed molecules are non-linear and/or have
strongly-resolved hyperfine structure, complications that make 
calculation of the collisional rate coefficients more difficult
and limit our interpretative abilities.

The calculation described here has only a very limited range of validity for 
CO whose small dipole moment (0.11 Debye) leads to 
A$_{10} \approx 7 \times 10^{-8}$\ps, some 600  times smaller than 
for  \hcop,  and it is not interesting.  Even far below the levels of
detectability, CO is already in a regime where the brightness of
its J=1-0 transition is proportional to column density and largely 
independent of density \citep{Lis07CO}.  Somewhat like C I, the 
excitation temperature of the J=1-0 line of CO is sensitive to the
thermal pressure of \HH, not the \HH\ density alone, when CO is 
strongly sub-thermally excited \citep{LisPet12}.

The calculation described here is also valid  over an interesting range
of parameter space for CH whose J=1-0 transition lies much higher near 
530 GHz, but CH can also be treated well as a two-level atom under
most circumstances in which it would be expected to have substantial
abundances.  Perhaps a  more interesting case would be CH\p\ that is 
produced and observed in shocked regions at higher temperature 
\citep{GodFal+12}.

\subsection{Distinguishing density and abundance variations}

The range of possible behaviour embodied in Eq. 9 and 11 for the line
of sight integral is richer and more complex than the case of constant 
density and temperature used in the comparison with LVG calculations 
in the previous Section. However, to the extent that different species 
observed over the same cloud sample more or less the same material and 
are subject to the same run of density n(H) at any position, their 
limiting brightness distributions will mainly reflect their abundances: 
If one species is more spatially confined than another, that 
is because of its chemistry -- the spatial distribution of its
abundance and column density -- not because it is more poorly excited.  
When the ratio of brightnesses of two species changes from position to 
position that mainly reflects variation in their relative abundances. 

In the weak excitation limit, the ratio of brightnesses of two lines of 
the same molecule given by Eq. 9 is nominally independent of the density, 
or even the variation of density along the line of sight, in even the 
(supposedly) most density-sensitive species.  If the ratio changes from 
position to position, the change can only arise from variations in the 
temperature or ionization fraction in the host gas.  For excitation by 
neutral particles, the brightness of the 
1-0 line of \hcop\ at left in Figure 1 increases by 50\% when going 
from 10 to 20 K while the brightness of the 2-1 line doubles.
There is relatively little temperature variation for the J=1-0 
lines when excited by electrons, but rather more variation for the J=2-1
line of HNC.

\section{Summary}

In Section 2 we  re-derived expressions for the emergent brightness of 
mm-wave molecular rotational transitions in the limit where the overall 
collisional excitation rates are small compared to the spontaneous emission 
rates, as defined by a suitably-defined maximum density n(H)$_{max}$.  
The emergent brightnesses of the lines of a species Y can be expressed  
as an emission measure - the line of sight integral of the product of two
number densities n(H) $\times$ n(Y) - with a constant of proportionality that 
depends on fundamental constants and an appropriately-defined kinetic 
temperature-dependent excitation rate coefficient.  The excitation rate
coefficient is found by summing over the rate constants for individual
upward transitions without solving the multi-level rate equations for the 
simultaneous equilibrium of all the individual states.

In this limit, the line brightnesses can be calculated without mention 
of the rotational excitation temperatures, optical depths, spontaneous 
emission rates or critical densities etc. that are usually cited as 
being important for the excitation and detectability of weak lines.  
This is a direct consequence of the fact that thermal energy injected 
into the molecular energy ladder by upward collisions emerges from the 
gas without being reabsorbed into 
the thermal energy reservoir by collisional de-excitation, even after 
repeated scatterings.  The only condition necessary for the detectability 
of a line is that the carrier molecule can channel enough of the 
ambient thermal energy into the line.  

In Section 3 we evaluated the formalism for the low-lying transitions of
the three commonly-observed molecules \hcop, HNC and CS.  To test the 
extent of the weak collisional excitation limit and the bulk excitation 
calculation we compared the results of the closed-form expressions for 
the line brightness with the results of LVG escape probability solutions 
of the multi-level equilibrium.  Using the limiting form for the escape 
probability $\beta \rightarrow 1/\tau$ at high optical depth, we derived
formal  limits on the product n(H)dN(Y)/dv that are directly equivalent
to limits on the emergent line brightness.   The brightness limits
are expressed in a simple form that has little reference to the 
underlying formalism or even the structure of the molecule in question.  
The closed form and numerical solutions diverge in expected ways, allowing
us to set the numerical value of the sole free parameter of the formalism, 
which defines the numerical meaning of ``weak'' excitation.

We tabulated parameters for \hcop, HNC and CS by limiting the divergence 
between the closed form and LVG approximations (both are approximations
of different kinds) to 20\% of the predicted line brightness in the most 
extreme case.  Limits on the allowable densities n(H)$_{max}$ in Tables 2 and 
3 are perhaps surprisingly high, typically $10^4 \pccc$ for excitation 
by \HH\ and He in the J=1-0 lines of \hcop, HNC, although 4-5 times smaller 
for CS.  Allowable densities are much larger for higher-lying lines and 
lower for excitation by electrons when  x$_e = 1.4\times 10^{-4}$
and the CO fraction in the gas is presumably small.  Limits on 
n(H)dN/dv and the limiting line brightnesses are shown in Table 4. 
The latter are 0.3 K for J=1-0 lines and 0.8 - 1.8 K for J=2-1.
Cases covered by the weak excitation limit may extend to up quite
high optical depth at appreciable density, for instance up to optical
depth $\tau_{01} = 10$ at n(H) $= 1000\pccc$ or $\tau_{01} = 100$ 
at n(H) $= 100\pccc$ for the calculations shown in Figure 1.

The full range of behaviour is implicitly contained in the defining
equations which relate the integrated  brightness of lines of species Y 
to the line of sight integral of n(H)n(Y) that is the molecular 
emission measure.  When the relative abundance n(Y)/n(H) is constant,
the integrated brightness is proportional to the line of sight integral 
of n(H)$^2$ which is susceptible to clumping and dominated by regions 
of high density.  When the density is constant, the integrated line 
brightness is proportional to n(H)N(Y) and perhaps weighted to regions
of higher kinetic temperature where the excitation rate is larger.
The typical LVG or escape probability calculation corresponds to the 
limiting case where everything inside the line of sight integral is 
constant,  so the expressions derived here can, under the proper
conditions,  replace grids 
of numerical solutions while making explicit the dependences on
the assumed density and temperature etc that are not necessarily 
exemplified in the numerical work.

\acknowledgments

  The National Radio Astronomy Observatory is operated by Associated
  Universities, Inc. under a cooperative agreement with the National Science 
  Foundation. IRAM is operated by CNRS (France), the MPG (Germany) and 
  the IGN (Spain). This work was in part supported by the CNRS program 
  ``Physique et Chimie du Milieu Interstellaire'' (PCMI).
  The authors thank Francois Lique for providing references,
  cross-sections and prompt answers to various questions, and Javier
  Goicoechea for comments on an early version of the present formalism.
  We thank an anonymous first referee for bringing to our attention the
  work by \cite{Pen75} and \cite{LinGol+77} and we thank the anonymous 
  second referee for helpful comments that cut the Gordian Knot of contention 
  with the first referee.  This work made use of the BASECOL collision browser
  \citep{DubAle13} and NASA ADS Abstract Service. 






\appendix

\section{Excitation of the 1-0 transition}


Tables 5 and 6 give Einstein A-coefficients for the upper levels J$_u$ and the
upward collision rate terms $f_0 \gamma_{0,J_u}/(1+p_\nu(T_{cmb})$ used
to compute the excitation of the J=1 level in Eq. 4.  Results are given for
for x$_e = 0$ and x$_e = 1.4\times 10^{-4}$.  Units are $10^{-10}$ cm$^3$\ps.

For excitation by \HH\ (Table 5), direct excitations into the J=1 level 
dominate for \TK\ $\la 10$ K but at 80 K they comprise only about 1/3 
of the total rate.  This is the origin of the temperature dependences shown
in Figures 1 - 3.  For electron excitation (Table 6) direct excitations into
the J=1 level dominate at all temperatures and the temperature dependence is
weaker, again as shown in Figures 1-3.

\begin{table*}
\caption[]{Einstein A-coefficients A$_{J_u,J_{u-1}}$ and upward collision rate
terms entering Eq. 4 with x$_e = 0$.}
{
\begin{tabular}{ccccccccc}
\hline
&&&&\hcop&&&&\\
\hline
$J_l-J_u$&A$_{J_u,J_{u-1}}$&  5 K  &  10 K   &  20 K   &  30 K   &  40 K   &  60 K  &  80 K  \\
       & \ps & &&&$10^{-10} {\rm cm}^3$ \ps  & & & \\
\hline
0-1 & 4.09E-05 & 1.2886 & 1.3877 & 1.4945 & 1.4630 & 1.4411 & 1.3937 & 1.3610\\
0-2 & 3.92E-04 & 0.3636 & 0.6125 & 1.0315 & 1.1406 & 1.2249 & 1.2249 & 1.2249\\
0-3 & 1.42E-03 & 0.0541 & 0.1803 & 0.6010 & 0.7944 & 0.9682 & 1.0443 & 1.1018\\
0-4 & 3.49E-03 & 0.0030 & 0.0249 & 0.2051 & 0.3606 & 0.5380 & 0.6821 & 0.8071\\
0-5 & 6.96E-03 & 0.0001 & 0.0020 & 0.0505 & 0.1285 & 0.2492 & 0.3913 & 0.5388\\
0-6 & 1.22E-02 & 0.0000 & 0.0002 & 0.0145 & 0.0526 & 0.1314 & 0.2536 & 0.4043\\
\hline
&&&&HNC&&&&\\
\hline
$J_l-J_u$&A$_{J_u,J_{u-1}}$&  5 K  &  10 K   &  20 K   &  30 K   &  40 K   &  60 K  &  80 K  \\
       & \ps & &&&$10^{-10} {\rm cm}^3$ \ps  & & & \\
\hline
0-1 & 2.60E-05 & 0.4009 & 0.5559 & 0.6532 & 0.7036 & 0.7506 & 0.8401 & 0.9050\\
0-2 & 2.49E-04 & 0.0895 & 0.2999 & 0.5245 & 0.6274 & 0.6793 & 0.7179 & 0.7351\\
0-3 & 9.02E-04 & 0.0020 & 0.0283 & 0.0978 & 0.1417 & 0.1789 & 0.2617 & 0.3303\\
0-4 & 2.22E-03 & 0.0000 & 0.0024 & 0.0254 & 0.0557 & 0.0852 & 0.1353 & 0.1678\\
0-5 & 4.43E-03 & 0.0000 & 0.0002 & 0.0046 & 0.0146 & 0.0266 & 0.0511 & 0.0736\\
0-6 & 7.77E-03 & 0.0000 & 0.0000 & 0.0005 & 0.0026 & 0.0060 & 0.0151 & 0.0252\\
\hline
&&&&CS&&&&\\
\hline
$J_l-J_u$&A$_{J_u,J_{u-1}}$&  5 K  &  10 K   &  20 K   &  30 K   &  40 K   &  60 K  &  80 K  \\
       & \ps & &&&$10^{-10} {\rm cm}^3$ \ps  & & & \\
\hline
0-1 & 1.69E-06 & 0.0404 & 0.1023 & 0.1028 & 0.1028 & 0.1020 & 0.1017 & 0.1018\\
0-2 & 1.62E-05 & 0.0221 & 0.0895 & 0.1206 & 0.1354 & 0.1433 & 0.1623 & 0.1813\\
0-3 & 5.87E-05 & 0.0019 & 0.0154 & 0.0316 & 0.0413 & 0.0474 & 0.0555 & 0.0609\\
0-4 & 1.44E-04 & 0.0004 & 0.0078 & 0.0266 & 0.0404 & 0.0500 & 0.0630 & 0.0730\\
0-5 & 2.88E-04 & 0.0000 & 0.0014 & 0.0091 & 0.0173 & 0.0242 & 0.0340 & 0.0404\\
0-6 & 5.06E-04 & 0.0000 & 0.0003 & 0.0038 & 0.0097 & 0.0158 & 0.0263 & 0.0344\\
\hline
\end{tabular}}
\\
\end{table*}

\begin{table*}
\caption[]{Einstein A-coefficients A$_{J_u,J_{u-1}}$ and upward collision rate 
terms entering Eq. 4 for x$_e = 1.4\times10^{-4}$.}
{
\begin{tabular}{cccccccccc}
\hline
&&&&\hcop&&&&\\
\hline
$J_l-J_u$&A$_{J_u,J_{u-1}}$&  5 K  &  10 K   &  20 K   &  30 K   &  40 K   &  60 K  &  80 K  \\
       & \ps & &&&$10^{-10} {\rm cm}^3$ \ps  & & & \\
\hline
0-1 & 4.09E-05 & 17.1290 & 18.9274 & 16.7029 & 14.7472 & 13.3841 & 11.6152 & 10.5138\\
0-2 & 3.92E-04 & 0.9935 & 2.2323 & 3.2158 & 3.3514 & 3.3553 & 3.1577 & 2.9868\\
0-3 & 1.42E-03 & 0.0627 & 0.2607 & 0.8078 & 1.0536 & 1.2462 & 1.3251 & 1.3722\\
0-4 & 3.49E-03 & 0.0032 & 0.0331 & 0.2550 & 0.4437 & 0.6408 & 0.8016 & 0.9307\\
0-5 & 6.96E-03 & 0.0001 & 0.0021 & 0.0527 & 0.1338 & 0.2570 & 0.4021 & 0.5511\\
0-6 & 1.22E-02 & 0.0000 & 0.0002 & 0.0146 & 0.0531 & 0.1323 & 0.2551 & 0.4061\\

\hline
&&&&HNC&&&&\\
\hline
$J_l-J_u$&A$_{J_u,J_{u-1}}$&  5 K  &  10 K   &  20 K   &  30 K   &  40 K   &  60 K  &  80 K  \\
       & \ps & &&&$10^{-10} {\rm cm}^3$ \ps  & & & \\
\hline
0-1 & 2.60E-05 & 4.8820 & 6.9694 & 7.9674 & 8.1270 & 8.1086 & 7.9518 & 7.7590\\
0-2 & 2.49E-04 & 0.0895 & 0.2999 & 0.5245 & 0.6274 & 0.6793 & 0.7179 & 0.7351\\
0-3 & 9.02E-04 & 0.0020 & 0.0283 & 0.0978 & 0.1417 & 0.1789 & 0.2617 & 0.3303\\
0-4 & 2.22E-03 & 0.0000 & 0.0024 & 0.0254 & 0.0557 & 0.0852 & 0.1353 & 0.1678\\
0-5 & 4.43E-03 & 0.0000 & 0.0002 & 0.0046 & 0.0146 & 0.0266 & 0.0511 & 0.0736\\
0-6 & 7.77E-03 & 0.0000 & 0.0000 & 0.0005 & 0.0026 & 0.0060 & 0.0151 & 0.0252\\
\hline
&&&&CS&&&&\\
\hline
$J_l-J_u$&A$_{J_u,J_{u-1}}$&  5 K  &  10 K   &  20 K   &  30 K   &  40 K   &  60 K  &  80 K  \\
       & \ps & &&&$10^{-10} {\rm cm}^3$ \ps  & & & \\
\hline
0-1 & 1.69E-06 & 1.1359 & 1.4180 & 1.4989 & 1.5084 & 1.5015 & 1.4756 & 1.4457\\
0-2 & 1.62E-05 & 0.0486 & 0.1366 & 0.1771 & 0.1920 & 0.1985 & 0.2137 & 0.2293\\
0-3 & 5.87E-05 & 0.0021 & 0.0162 & 0.0331 & 0.0429 & 0.0492 & 0.0572 & 0.0626\\
0-4 & 1.44E-04 & 0.0004 & 0.0078 & 0.0267 & 0.0404 & 0.0501 & 0.0630 & 0.0730\\
0-5 & 2.88E-04 & 0.0000 & 0.0014 & 0.0091 & 0.0173 & 0.0242 & 0.0340 & 0.0404\\
0-6 & 5.06E-04 & 0.0000 & 0.0003 & 0.0038 & 0.0097 & 0.0158 & 0.0263 & 0.0344\\
\hline
\end{tabular}}
\\
\end{table*}

\bibliographystyle{apj}

\begin{thebibliography}{39}
\expandafter\ifx\csname natexlab\endcsname\relax\def\natexlab#1{#1}\fi

\bibitem[{{Balakrishnan} {et~al.}(2002){Balakrishnan}, {Yan}, \&
  {Dalgarno}}]{BalYan+02}
{Balakrishnan}, N., {Yan}, M., \& {Dalgarno}, A. 2002, ApJ, 568, 443 

\bibitem[{{Balser}(2006)}]{Bal06}
{Balser}, D.~S. 2006, AJ, 132, 2326

\bibitem[{{Bhattacharyya} {et~al.}(1981){Bhattacharyya}, {Bhattacharyya}, \&
  {Narayan}}]{BhaBha+81}
{Bhattacharyya}, S.~S., {Bhattacharyya}, B., \& {Narayan}, M.~V. 1981, ApJ,
  247, 936

\bibitem[{{Buffa} {et~al.}(2009){Buffa}, {Dore}, \& {Meuwly}}]{BufDor+09}
{Buffa}, G., {Dore}, L., \& {Meuwly}, M. 2009, MNRAS., 397,
  1909

\bibitem[{{Castor}(1970)}]{Cas70}
{Castor}, J.~I. 1970, MNRAS, 149, 111

\bibitem[{{Dickinson} \& {Flower}(1981)}]{DicFlo81}
{Dickinson}, A.~S. \& {Flower}, D.~R. 1981, MNRAS, 196, 297

\bibitem[{{Dickinson} {et~al.}(1977){Dickinson}, {Phillips}, {Goldsmith},
  {Percival}, \& {Richards}}]{DicPhi+77}
{Dickinson}, A.~S., {Phillips}, T.~G., {Goldsmith}, P.~F., {Percival}, I.~C.,
  \& {Richards}, D. 1977, A\&A, 54, 645

\bibitem[{{Dubernet} {et~al.}(2013){Dubernet}, {Alexander}, {Ba},
  {Balakrishnan}, {Balanca}, {Ceccarelli}, {Cernicharo}, {Daniel}, {Dayou},
  {Doronin}, {Dumouchel}, {Faure}, {Feautrier}, {Flower}, {Grosjean}, 
  {Halvick}, {K{\l}os}, {Lique}, {McBane}, {Marinakis}, {Moreau}, {Moszynski},
  {Neufeld}, {Roueff}, {Schilke}, {Spielfiedel}, {Stancil}, {Stoecklin},
  {Tennyson}, {Yang}, {Vasserot}, \& {Wiesenfeld}}]{DubAle13}
{Dubernet}, M.-L., {Alexander}, M.~H., {Ba}, Y.~A., et al.  2013, A\&A, 553, A50

\bibitem[{{Dumouchel} {et~al.}(2010){Dumouchel}, {Faure}, \&
  {Lique}}]{DumFau+10}
{Dumouchel}, F., {Faure}, A., \& {Lique}, F. 2010, MNRAS,
  406, 2488

\bibitem[{{Dumouchel} {et~al.}(2011){Dumouchel}, {K{\l}os}, \&
  {Lique}}]{DumKlo+11}
{Dumouchel}, F., {K{\l}os}, J., \& {Lique}, F. 2011, Physical Chemistry
  Chemical Physics (Incorporating Faraday Transactions), 13, 8204

\bibitem[{{Faure} {et~al.}(2007{\natexlab{a}}){Faure}, {Tennyson}, {Varambhia},
  {Kokoouline}, {Greene}, \& {Stoecklin}}]{FauTen+07}
{Faure}, A., {Tennyson}, J., {Varambhia}, H.~N.,
 et al. 2007{\natexlab{a}}, in Molecules in Space and
  Laboratory, ed. J. L. Lemaire, \& F. Combes

\bibitem[{{Faure} {et~al.}(2007{\natexlab{b}}){Faure}, {Varambhia},
  {Stoecklin}, \& {Tennyson}}]{FauVar+07}
{Faure}, A., {Varambhia}, H.~N., {Stoecklin}, T., \& {Tennyson}, J.
  2007{\natexlab{b}}, MNRAS, 382, 840

\bibitem[{{Flower}(1999)}]{Flo99}
{Flower}, D.~R. 1999, MNRAS, 305, 651

\bibitem[{{Godard} {et~al.}(2012){Godard}, {Falgarone}, {Gerin}, {Lis}, {De
  Luca}, {Black}, {Goicoechea}, {Cernicharo}, {Neufeld}, {Menten}, \&
  {Emprechtinger}}]{GodFal+12}
{Godard}, B., {Falgarone}, E., {Gerin}, M., et al. A\&A, 540, A87

\bibitem[{{Goldreich} \& {Kwan}(1974)}]{GolKwa74}
{Goldreich}, P. \& {Kwan}, J. 1974, ApJ, 189, 441

\bibitem[{{Goldsmith} {et~al.}(2012){Goldsmith}, {Langer}, {Pineda}, \&
  {Velusamy}}]{GolLan+12}
{Goldsmith}, P.~F., {Langer}, W.~D., {Pineda}, J.~L., \& {Velusamy}, T. 2012,
  Astrophys. J., Suppl. Ser., 203, 13

\bibitem[{{Helfer} \& {Blitz}(1997)}]{HelBli97}
{Helfer}, T.~T. \& {Blitz}, L. 1997, ApJ, 478, 233

\bibitem[{{Levrier} {et~al.}(2012){Levrier}, {Le Petit}, {Hennebelle},
  {Lesaffre}, {Gerin}, \& {Falgarone}}]{LevLeP+12}
{Levrier}, F., {Le Petit}, F., {Hennebelle}, P., et al. 2012, A\&A, 544, A22

\bibitem[{{Linke} {et~al.}(1977){Linke}, {Goldsmith}, {Wannier}, {Wilson}, \&
  {Penzias}}]{LinGol+77}
{Linke}, R.~A., {Goldsmith}, P.~F., {Wannier}, P.~G., {Wilson}, R.~W., \&
  {Penzias}, A.~A. 1977, ApJ, 214, 50

\bibitem[{{Lique} {et~al.}(2006){Lique}, {Spielfiedel}, \&
  {Cernicharo}}]{LiqSpi+06}
{Lique}, F., {Spielfiedel}, A., \& {Cernicharo}, J. 2006, A\&A, 451, 1125

\bibitem[{{Liszt}(1995)}]{Lis95}
{Liszt}, H.~S. 1995, ApJ, 442, 163

\bibitem[{{Liszt}(2006)}]{Lis06}
---. 2006, A\&A, 458, 507

\bibitem[{{Liszt}(2007)}]{Lis07CO}
---. 2007, A\&A, 476, 291

\bibitem[{{Liszt}(2012)}]{Lis12}
---. 2012, A\&A, 538, A27

\bibitem[{{Liszt} \& {Pety}(2012)}]{LisPet12}
{Liszt}, H.~S. \& {Pety}, J. 2012, A\&A, 541, A58

\bibitem[{{Lucas} \& {Liszt}(1996)}]{LucLis96}
{Lucas}, R. \& {Liszt}, H.~S. 1996, A\&A, 307, 237

\bibitem[{{Mangum} \& {Shirley}(2015)}]{ManShi15}
{Mangum}, J.~G. \& {Shirley}, Y.~L. 2015, PASP, 127, 266

\bibitem[{{Neufeld} \& {Dalgarno}(1989)}]{NeuDal89}
{Neufeld}, D.~A. \& {Dalgarno}, A. 1989, Phys. Rev. A, 40, 633


\bibitem[{{Penzias}(1975)}]{Pen75}
{Penzias}, A.~A. 1975, in Atomic and Molecular Physics and the Interstellar
  Matter, ed. R.~{Balian}, P.~{Encrenaz}, \& J.~{Lequeux}
(Les Houches session 26; Amsterdam: North-Holland), 373--408

\bibitem[{{Shepler} {et~al.}(2007){Shepler}, {Yang}, {Dhilip Kumar}, {Stancil},
  {Bowman}, {Balakrishnan}, {Zhang}, {Bodo}, \& {Dalgarno}}]{SheYan+07}
{Shepler}, B.~C., {Yang}, B.~H., {Dhilip Kumar}, T.~J., et al.
  A. 2007, A\&A, 475, 15

\bibitem[{{Shirley}(2015)}]{Shi15}
{Shirley}, Y.~L. 2015, PASP, 127, 299

\bibitem[{{Snell} {et~al.}(2000){Snell}, {Howe}, {Ashby}, {Bergin}, {Chin},
  {Erickson}, {Goldsmith}, {Harwit}, {Kleiner}, {Koch}, {Neufeld}, {Patten},
  {Plume}, {Schieder}, {Stauffer}, {Tolls}, {Wang}, {Winnewisser}, {Zhang}, \&
  {Melnick}}]{SneHow+00}
{Snell}, R.~L., {Howe}, J.~E., {Ashby}, M.~L.~N., et al. 2000, ApJ, 539, L93

\bibitem[{{Spitzer}(1978)}]{Spi78}
{Spitzer}, L. 1978, Physical processes in the interstellar medium (New York
  Wiley-Interscience, 1978. 333 p.)

\bibitem[{{van der Tak} {et~al.}(2007){van der Tak}, {Black}, {Sch{\"o}ier},
  {Jansen}, \& {van Dishoeck}}]{RADEX}
{van der Tak}, F.~F.~S., {Black}, J.~H., {Sch{\"o}ier}, F.~L., {Jansen}, D.~J.,
  \& {van Dishoeck}, E.~F. 2007, A\&A, 468, 627

\bibitem[{{Varambhia} {et~al.}(2010){Varambhia}, {Faure}, {Graupner}, {Field},
  \& {Tennyson}}]{VarFau+10}
{Varambhia}, H.~N., {Faure}, A., {Graupner}, K., {Field}, T.~A., \& {Tennyson},
  J. 2010, MNRAS, 403, 1409

\bibitem[{{Wannier} {et~al.}(1991){Wannier}, {Pagani}, {Kuiper}, {Frerking},
  {Gulkis}, {Encrenaz}, {Pickett}, {Lecacheux}, \& {Wilson}}]{WanPag+91}
{Wannier}, P.~G., {Pagani}, L., {Kuiper}, T.~B.~H., et al. 1991, ApJ, 377, 171

\bibitem[{{Yang} {et~al.}(2013){Yang}, {Stancil}, {Balakrishnan}, {Forrey}, \&
  {Bowman}}]{YanSta+13}
{Yang}, B., {Stancil}, P.~C., {Balakrishnan}, N., {Forrey}, R.~C., \& {Bowman},
  J.~M. 2013, ApJ, 771, 49

\bibitem[{{Yazidi} {et~al.}(2014){Yazidi}, {Ben Abdallah}, \&
  {Lique}}]{YazBen+14}
{Yazidi}, O., {Ben Abdallah}, D., \& {Lique}, F. 2014, MNRAS, 441, 664

\end{thebibliography}

\end{document}